\documentclass[prb,twocolumn,aps,showpacs]{revtex4}
\usepackage{graphicx}

\begin{document}
\title{
Strain-induced interface reconstruction in epitaxial heterostructures
}
\author{N. Lazarides$^{1,2}$, V. Paltoglou$^{1,2}$, P. Maniadis$^{1,2}$, 
G. P. Tsironis$^{1,2}$, and C. Panagopoulos$^{1,2,3}$,
}
\affiliation{
$^{1}$Department of Physics, University of Crete, P. O. Box 2208, 71003 Heraklion, 
      Greece \\
$^{2}$Institute of Electronic Structure and Laser, 
      Foundation for Research and Technology-Hellas, P.O. Box 1527, 71110 Heraklion,
      Greece \\
$^{3}$Division of Physics and Applied Physics, Nanyang Technological University,
      637371, Singapore
}

\date{\today}
\begin{abstract}
We investigate in the framework of Landau theory the distortion of the strain fields 
at the interface of two dissimilar ferroelastic oxides that undergo a structural 
cubic-to-tetragonal phase transition.
Simple analytical solutions are derived for the dilatational and the deviatoric 
strains that are globally valid over the whole of the heterostructure.
The solutions reveal that the dilatational strain exhibits compression close to the 
interface which may in turn affect the electronic properties in that region. 
\end{abstract}

\maketitle

\section{Introduction
}
Recent discoveries in material science related to several 
unexpected properties of epitaxial heterostructures made of different 
transition metal oxide (TMO) materials,
bring in the forefront of interest the problem of interface reconstruction 
through the developement of spontaneous strain at the interface
\cite{Ohtomo2002,Ohtomo2004}.
Lattice distortion close to the interface is known to result in
charge redistribution that leads to the formation of a two-dimensional electron 
gas (2DEG) and metallicity in that region 
\cite{Shibuya2004,Okamoto2006,Ishida2008,Wong2010}.
Most of the TMOs of interest are ferroelastics that undergo structural
transitions \cite{Salje1990} from a cubic/pseudocubic 
to a lower symmetry phase with decreasing temperature. 
Notably, heterostructures containing strontium
titanate (SrTiO$_3$), a band-insulator oxide undergoing a cubic-to-tetragonal 
(CTT) structural transition at $T_s \sim 105~^oK$, 
exhibit extraordinary interfacial properties below $T_s$; 
metallicity \cite{Ohtomo2002,Shibuya2004,Seo2007,Wong2010},
superconductivity \cite{Biscaras2010}, and nonlinear Hall effect \cite{Kim2010}. 
Moreover, in LaTiO$_3$/SrTiO$_3$ and LaAlO$_3$/SrTiO$_3$ heterostructures, 
the structural transition of SrTiO$_3$ causes the overlayers to stabilize in 
a tetragonal phase with an in-plane lattice constant almost equal to that of
SrTiO$_3$ close to the interface \cite{Kim2003,Wu2011}.

It has been discussed in the past that the electromagnetic properties of
TMOs couple to the elastic degrees of freedom
\cite{Ahn2004,Bishop2008,Maniadis2008}.
The effect of tensile and compressive strains to the electronic conduction 
properties at the interface of TMO heterostructures has already been addressed 
experimentally \cite{Bark2011,JWSeo2010}.
Furthermore, strong polarization enhancement in ferroelectric TMO superlattices 
driven by interfacial strain has been unambigiously observed \cite{Lee2005}. 
In the present work we apply continuous elasticity theory through a 
Ginzburg-Landau description in terms of the strain tensor components to a 
heterostructure.
Based solely on symmetry considerations, Ginzburg-Landau theory can provide
a reliable description of the equilibrium behavior of a system near a phase 
transition.
It has been recently used to show theoretically the emergence of a multiferroic
state of a EuTiO$_3$ film on 
(LaAlO$_3)_{0.29}$–(SrAl$_{1/2}$Ta$_{1/2}$O$_3)_{0.71}$ 
(LSAT) substrate \cite{Lee2010},
%%%% LSAT (LaAlO$_3$)$_0.29$–(SrAl$_{1/2}$Ta$_{1/2}O$_$3)$_{0.71}$
to provide a physical understanding of the strain-induced metal-insulator
phase coexistence in manganites \cite{Ahn2004}, 
and to explain phase separation between metallic ferromagnetic and insulating
charge-modulated phases \cite{Milward2005}.

We investigate the interfacial effects on the strain-state of a bilayer 
heterostructure, composed of dissimilar TMOs that join at a single planar
interface
and propose a strain-based mechanism that may help understand the formation
of a 2DEG.
In particular, we obtain approximate analytical solutions for the dilatational and 
the deviatoric strain fields in the bilayer, that exhibit spatial variation due
to breaking of the uniformity.
Notably, the dilatational strain field exhibits a well-defined minimum at the 
interface corresponding to local compression \cite{Wong2010}.
We argue that the suppression of the dilatational strain field in the interfacial 
region may encourage the formation of a 2DEG.
The proposed strain-based mechanism does not exclude other possible mechanisms, 
like, e.g., the orbital and/or the electronic reconstruction mechanisms
\cite{Chakhalian2007,Okamoto2004}. 

\section{Ginzburg-Landau theory and equations of motion
}
In the Lagrangian description of elasticity the symmetric strain tensor is defined 
as $\epsilon_{ij} =  \left\{ u_{i,j} + u_{j,i} \right\}/2$ ($i,j =x,y,z$),
where $u_{i,j}$ is the $j-$th derivative of the $i-$th component of the displacement 
vector $\bf u$ of a material point relative to its position in the parent phase.
The six symmetry adapted strains for the CTT structural transition are defined as
\cite{Jacobs2003}
\begin{eqnarray}
  \label{2.1}
  e_1 = u_{x,x} +u_{y,y} +u_{z,z} ,~~ 
  e_2 = \frac{1}{2} (u_{x,x} -u_{y,y}) \\
  \label{2.3}
  e_3 = \frac{1}{2\sqrt{3}} ( u_{x,x} +u_{y,y} -2 u_{z,z} ) ,~~ 
  e_4 = \frac{1}{2} ( u_{y,z} + u_{z,y} ) , 
\end{eqnarray}
while $e_5$ and $e_6$ are given by $e_4$ with cyclic permutation of the indices.
The deviatoric strains $e_2$ and $e_3$ form the two-component order parameter (OP)
of the CTT transition. 
Both the OP and the non-OP strains are coordinate-independent in the uniform 
product (tetragonal) phase in static equilibrium, with the latter customarily being 
set to zero.  In a TMO heterostructure, where the uniformity of the product phase 
is broken due to the interface, all $e_i$'s vary spatially; 
in that case, their second derivatives are linked through
compatibility relations \cite{Rasmussen2001}.
In a non-uniform state, the non-OP strains cannot be all set to zero. 
Specifically, in TMO heterostructures the dilatational strain $e_1$, 
which is concommitant to $e_3$ \cite{Kosogor2011,Chernenko1996}, exhibits 
measurable compression indicating its importance in their structural 
properties \cite{Wong2010}. 

In ferroelasticity theory, the strain energy density ${\cal F}$ of a material undergoing
a CTT structural transition is expanded in powers of the invariants of the strain
tensor and their products around the energy of the parent phase 
\cite{Kosogor2011,Chernenko1996,Barsch1984,Gomonaj1994,Gomonaj1996,Rasmussen2001}. 
Thus, the functional ${\cal F}$ is expressed solely in terms of the $e_i$'s
and their spatial derivatives. Guided by previous works we adopt a functional 
${\cal F}$ of the form 
\begin{eqnarray}
 \label{8}
  {\cal F} = 
      \frac{c_1}{2} e_1^2 +\frac{c_2}{2} (e_2^2 +e_3^2)
       +\frac{c_3}{2} (e_4^2 +e_5^2 +e_6^2) 
       \nonumber \\
       + \frac{d_1}{2} \left( {\bf \nabla} e_1 \right)^2
       + \frac{d_2}{2} \left[ \left( {\bf \nabla} e_2 \right)^2
                             +\left( {\bf \nabla} e_3 \right)^2 \right] 
      \nonumber \\ 
      +\frac{a_1}{2} e_1^3 +\frac{a_2}{2} e_1 (e_2^2 +e_3^2)
      +\frac{a_4}{3} e_3 (e_3^2 -3e_2^2) 
      +\frac{b_1}{4} e_1^4 
       \nonumber \\
     +\frac{b_2}{2} e_1^2 (e_2^2 +e_3^2)
     +\frac{b_4}{4} (e_2^2 +e_3^2)^2
     +\frac{b_7}{2} e_1 e_3 (e_3^2 -3e_2^2) ,  
\end{eqnarray}
where the Ginzburg-Landau coefficients $a_1,a_2,a_4$, $b_1,b_2,b_4,b_7$, and 
$c_1, c_2, c_3$ are related to the second-, third-, and fourth-order elastic 
coefficients of the parent phase, respectively, through (in Voigt notation)
\cite{Liakos1982}
\begin{eqnarray}
\label{9}
  a_1 &=& \frac{1}{27} \left( C_{111} +6 C_{112} +2 C_{123} \right) \nonumber \\
  a_2 &=& \frac{2}{3} ( C_{111} -C_{123} ) \nonumber \\
  a_4 &=& -\frac{1}{\sqrt{3}} \left( C_{111} -3 C_{112} +2 C_{123} \right) \nonumber \\
  b_1 &=& \frac{1}{162} ( C_{1111} +8 C_{1112} +6 C_{1122} +12 C_{1123} )  \nonumber \\
  b_2 &=& \frac{1}{9} ( C_{1111} +2 C_{1112} -3 C_{1123} ) \\
  b_4 &=& \frac{1}{3} ( C_{1111} -4 C_{1112} +3 C_{1122} ) \nonumber \\
  b_7 &=& -2\frac{\sqrt{3}}{27} ( C_{1111} -C_{1112} -3 C_{1122} +3 C_{1123} ) \nonumber \\
  c_1 &=& \frac{1}{3} ( C_{11} +2 C_{12} ) \nonumber \\
  c_2 &=& 2 ( C_{11} - C_{12} ) ,  \nonumber
\end{eqnarray}
while $d_1,d_2$ are two independent strain-gradient coefficients.
In accordance with common principles of Landau theory, the critical temperature 
dependence of the $c_2$ elastic constant, $c_2 \propto (T-T_s)$, is supposed to 
be true close to the transition point.
%%%%%%%%%%%%%%%%%%%%%%%%%%%%%FIG.01%%%%%%%%%%%%%%%%%%%%%%%%%%%%%%%%%%%%%
\begin{figure}[t!]
\includegraphics[scale=0.56,angle=0]{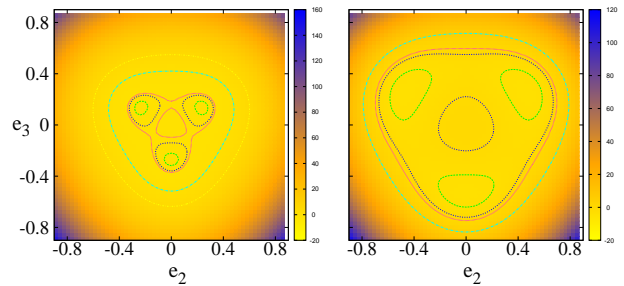}
\caption{
(color online)
Strain energy density lanscape on the $e_2 - e_3$ plane for 
$a_4 =25.4$, $b_4 =225$, and $c_2 =-10$ (left); $-50$ (right),
exhibiting the familiar pattern of three degenerate minima. 
}
\end{figure}
%%%%%%%%%%%%%%%%%%%%%%%%%%%%%%%%%%%%%%%%%%%%%%%%%%%%%%%%%%%%%%%%%%%%%%%%

In a single material at static equilibrium, the spatially homogeneous strains 
in the product phase are the lowest energy solutions of the
conditions ${\partial {\cal F} }/{\partial e_i} =0$. 
Neglecting the non-OP terms, the energy density landscape on the $e_2 -e_3$
plane exhibits the familiar pattern of three degenerate minima corresponding
to three different variants in the tetragonal phase (Fig. 1).
Notably, a non-zero $e_1$ preserves the energetic degeneracy of the three
variants. We are particularly interested in the variant having $e_2=0$; 
this is because TMO heterostructures are usually grown along the $z-$direction
and both materials go into a $c-$tetragonal phase at low temperatures
\cite{Kim2003,Wu2011}. 
For $e_2=0$, the energy landscape on the $e_1 -e_3$ plane shown in Fig. 2 
exhibits significant qualitative differences for different $a_2$ values.
Specifically, the strain $e_1$ varies from positive (i.e., expansional)
to negative (i.e., compressional) with decreasing the magnitude of $a_2$.
A small $a_2$ absolute value is however expected from the principle 
$e_1 << e_2, e_3$ for all martensitic transformations \cite{Gomonaj1994,Gomonaj1996}. 
Moreover, a small $a_2$ leads to negative $e_1$ and $e_3$,
in accordance with the empirical principle for the ferroelastic transitions 
of close-packed solids, i.e., that cooling of the solid is usually accompanied 
by a decrease of volume.   
The dependence of the strains on $c_2$ for all three variants is shown in Fig. 3
for two different values of $a_2$. 
We later refer to the two materials forming the bilayer heterostructure,
which occupy the  regions $z<0$ and $z>0$, as the left ($L$) and the right ($R$)
material, respectively.
The Ginzburg-Landau parameters used in Figs. 1-3 are those given for the left
material in Table I, and they have been calculated from the corresponding 
elastic coefficients through Eqs. (\ref{9}). 
The second- and third-order elastic coefficients of the left material
are those reported for SrTiO$_3$ \cite{Bell1963,Beattie1971}, 
while for the fourth-order ones a reasonable choise was made (Table I). 
Note that for the parameter $a_2$, which can be treated as a phenomenological 
one, we have also used values that are smaller than the one given in Table 
I for the left material (i.e., $a_2^L=-31$).
%%%%%%%%%%%%%%%%%%%%%%%%%%%%%FIG.02%%%%%%%%%%%%%%%%%%%%%%%%%%%%%%%%%%%%%
\begin{figure}[t!]
\includegraphics[scale=0.56,angle=0]{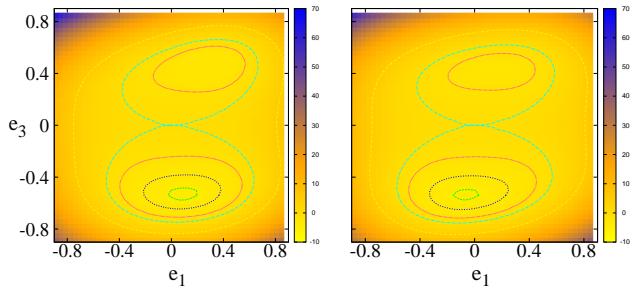}
\caption{
(color online)
Strain energy density lanscape on the $e_1 - e_3$ plane for
$c_2 =-50$, and $a_2 =-31$ (left); $-10$ (right).
The other Ginzburg-Landau parameters are those for the left material given
in Table I.
}
\end{figure}
%%%%%%%%%%%%%%%%%%%%%%%%%%%%%%%%%%%%%%%%%%%%%%%%%%%%%%%%%%%%%%%%%%%%%%%%

The dynamics of the displacements is governed by the Euler-Lagrange equations 
\begin{equation}
\label{10}
 \rho_0 \ddot{u}_i = \sigma_{ik,k} +\sigma_{ik,k}' ,
\end{equation}
where 
\begin{equation}
\label{11}
  \sigma_{ik} \equiv \frac{\partial {\cal F}}{\partial u_{i,k}}, \qquad
  \sigma_{ik}' \equiv \frac{\partial {\cal R}}{\partial \dot{u}_{i,k}} ,
\end{equation}
are the strain tensor and the dissipative strain tensor, respectively,
$\rho_0$ is the density in the parent phase,
and ${\cal R}$ is the the Rayleigh dissipation function
\begin{equation}
 \label{12}
   {\cal R} =\frac{1}{2} c_1' \dot{e}_1^2 +\frac{1}{2} c_2' (\dot{e}_2^2 +\dot{e}_3^2)
       +\frac{1}{2} c_3' (\dot{e}_4^2 +\dot{e}_5^2 +\dot{e}_6^2) . 
\end{equation}
Then, from Eq. (\ref{10}) we get 
\begin{equation}
 \label{20}
  \rho_0 \ddot{u}_i = \partial_i \Phi_i +\frac{1}{2} ( c_3 H_i +c_3' \dot{H}_i ) , 
\end{equation}
where 
\begin{equation}
 \label{13}
   H_x =e_{6,y} +e_{5,z}, ~H_y = e_{4,z} +e_{6,x},~H_z = e_{4,y} +e_{5,x} ,
\end{equation}
and
\begin{eqnarray}
 \label{21}
  \Phi_i=-\nabla^2 G_i +W_i +R_i +\dot{W}_i' , \\ 
 \label{21a}
  W_i = c_1 e_1 +\frac{c_2}{2} \left( q e_2 +\frac{1}{\sqrt{3}} e_3 \right) , 
  W_z = c_1 e_1 -\frac{c_2}{\sqrt{3}} e_3 , 
\end{eqnarray}
with $q=+1 ~(-1)$ for $i=x ~(y)$.
%%%%%%%%%%%%%%%%%%%%%%%%%%%%%FIG.03%%%%%%%%%%%%%%%%%%%%%%%%%%%%%%%%%%%%%
\begin{figure}[t!]
\includegraphics[scale=0.38,angle=0]{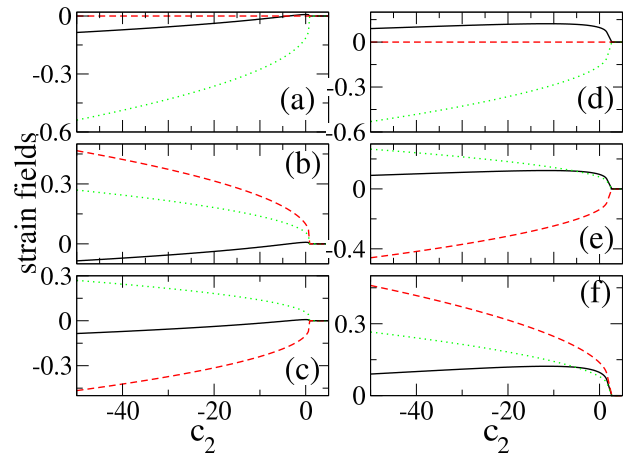}
\caption{
(color online)
The strains $e_1$ (black-solid), $e_2$ (red-dashed), $e_3$ (green-dotted),
as a function of $c_2$ for the three variants, 
for $a_2 =-7$ (left panels); $-31$ (right panels).
The other Ginzburg-Landau parameters are those for the left material given
in Table I.
}
\end{figure}
%%%%%%%%%%%%%%%%%%%%%%%%%%%%%%%%%%%%%%%%%%%%%%%%%%%%%%%%%%%%%%%%%%%%%%%%
The functions $W_i'$ and $G_i$ have the same form with that of the $W_i$'s, 
with the obvious change $c_j \rightarrow c_j'$ and $c_j \rightarrow d_j$,
respectivelly ($j=1,2$), and the $R_i$'s are lengthy nonlinear functions 
of $e_1$, $e_2$, and $e_3$, which are given in Appendix A.

\section{Approximate solutions and interface reconstruction
}
In order to separate the interfacial effects on the strain-state of the 
bilayer heterostructure (from those originating from external boundaries, 
domain walls, dislocations, etc.), we consider two monodomain, semi-infinite
TMOs joined along a chemically abrupt, planar interface at $z=0$. 
Eqs. (\ref{20}) could be simplified in a strict way,
since at low temperatures the strains depend on one coordinate only and $e_2=0$ 
\cite{Chernenko2006,Chernenko2006b}.
However, for non-zero $e_1$ the simpification of Eqs. (\ref{20}) following the 
strict way is a non-trivial task, which makes preferable the use of a simple 
{\em ansatz} for the displacements. This ansatz assures that the strains depend
only on the $z-$coordinate and that $e_2$, as well as the small strains
$e_4$, $e_5$, $e_6$, are identically zero.
Assume that the strains exhibit a relatively strong $z-$coordinate dependence
in the proximity of the interface,
while they attain their static equilibium values for large enough $|z|$.
This approximation seems well-suited for heterostuctures composed of TMOs with 
small lattice mismatch (i.e., LaTiO$_3$/SrTiO$_3$).  
Indeed, both experimental observations \cite{Helvoort2005} (discussed below)
and first-principles calculations \cite{Okamoto2006}
indicate that strain inhomogeneity and lattice deformation occur within
a few layers near the interface.
Thus, for practical purposes, it is sufficient for the two layers of the
heterostructure to be thick enough for the deformation to vanish relatively
far from the interface. The choise of semi-infinite layers was made only for 
mathematical convinience.
 
We then introduce the {\em ansatz}
\begin{eqnarray}
  \label{23}
  u_x =-\frac{a}{2} x, ~~~u_y =-\frac{a}{2} y, ~~~u_z=b z +f(z) ,
\end{eqnarray}
where $f(z)$ is a yet unknown function, and
\begin{equation}
\label{24}
   a=-\frac{2}{3} ( e_{10} +\sqrt{3} e_{30} ) , \qquad
   b=+\frac{1}{3} ( e_{10} -2\sqrt{3} e_{30} ) ,
\end{equation} 
with $e_{10}$ and $e_{30}$ being the values of $e_{1}$ and $e_{3}$, respectively,
far from the interface. The non-zero strains are then
\begin{equation}
  \label{26}
  e_1 =e_{10} +f'(z) , \qquad e_3 =e_{30} -\frac{1}{\sqrt{3}} f'(z) ,
\end{equation}
where the prime denotes differentiation with respect to $z$.
Substitution of Eq. (\ref{26}) into Eqs. (\ref{20}) results, in the static limit, 
in the equation
\begin{equation}
\label{26a}
  \left( d_1 +\frac{d_2}{3} \right) G''' = 
  \left( c_1 +\frac{c_2}{3} \right) G' + R_z' (G) ,
\end{equation}
where $G = G(z) \equiv f'(z)$, and $R_z$ with $e_2=0$ is
\begin{eqnarray}
 \label{22c}
  R_z = \frac{3 a_1}{2} e_1^2 +b_1 e_1^3 
       +\left( \frac{a_2}{2} -\frac{a_4}{\sqrt{3}} \right) e_3^2
       +\left( \frac{b_7}{2} -\frac{b_4}{\sqrt{3}} \right) e_3^3
           \nonumber \\
       -\frac{1}{\sqrt{3}} e_1 e_3 \left[
           a_2 +b_2 e_1 +3 \left( \frac{b_7}{2} -\frac{b_2}{\sqrt{3}} \right) e_3 
                          \right] ,~~~~~
\end{eqnarray}
where $e_1$ and $e_3$ are meant to be expressed in terms of $f$ through Eq. (\ref{26}).
After rearrangement, Eq. (\ref{26a}) becomes 
\begin{equation}
  \label{27}
  G''' =G' ( \tilde{\kappa} +3\tilde{\lambda} G +6\tilde{\mu} G^2 ), 
\end{equation}
where
\begin{equation}
\label{28a}
  \tilde{\kappa}=\frac{a_{1z} +\kappa}{d_{1z}},
  ~~\tilde{\lambda}= \frac{\lambda}{3 d_{1z}},
  ~~\tilde{\mu}=\frac{\mu}{6 d_{1z}} ,
\end{equation} 
with
\begin{eqnarray}
\label{28b}
  a_{1z} =c_1 +\frac{1}{3} c_2 =C_{11} , \qquad d_{1z} =d_1 +\frac{1}{3} d_2 , 
\end{eqnarray}
and 
\begin{eqnarray}
 \kappa&=&b C_{111} -3 a C_{112} +\frac{3 a^2}{4} ( C_{1122} +C_{1123} ) 
       -3ab C_{1112}  \nonumber \\
       &&+\frac{b^2}{2} C_{1111} , \nonumber \\
\label{28c}
 \lambda&=&c_{111} +( b C_{1111} -a C_{1112} ) , \\
 \mu&=&\frac{1}{2} C_{1111} . \nonumber
\end{eqnarray}

Eq. (\ref{27}) can be reduced to a quadrature that has the analytic solution
\begin{equation}
\label{30}
  G^{\pm} = 4\tilde{\kappa} \left( e^{\pm \sqrt{\tilde{\kappa}}(z \mp z_0)}
              +\Delta e^{\mp \sqrt{\tilde{\kappa}}(z \mp z_0)} +\Delta_1 \right)^{-1} ,  
\end{equation}
where $-\Delta\equiv 4 \tilde{\kappa} \tilde{\mu} -\tilde{\lambda}^2$, 
$\Delta_1 =-2 \tilde{\lambda}$, and $z_0$ is a constant of integration. 
Integration of $G^{\pm}$ gives 
\begin{equation}
\label{31}
  f^{\pm} = \mp \frac{2}{\sqrt{\tilde{\mu}}} \tanh^{-1} \left( 
     \frac{e^{\pm \sqrt{\tilde{\kappa}}(z \mp z_0)} -\tilde{\lambda}}
          {2 \sqrt{\tilde{\kappa} \tilde{\mu}} } \right) +C^{\pm} , 
\end{equation}
where $C^{\pm}$ are constants of integration.

The displacements and the strains in each material of the bilayer
can be written in terms of $f^\pm$ and $G^\pm$ from Eqs. (\ref{23}) and (\ref{26}), 
respectively.
%%%%%%%%%%%%%%%%%%%%%%%%%%%%%FIG.04%%%%%%%%%%%%%%%%%%%%%%%%%%%%%%%%%%%%%
\begin{figure}[t!]
\includegraphics[scale=0.38,angle=0]{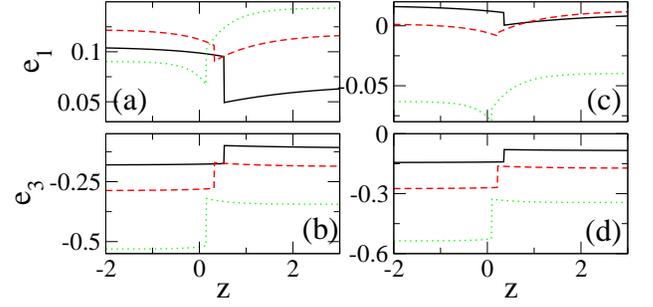}
\caption{(color online)
The strains $e_1$ and $e_3$ as a function of $z$ for 
$c_2^L=-1$, $c_2^R=-0.76$ (black-solid), $c_2^L=-10$, $c_2^R=-7.56$ (red-dashed);
$c_2^L=-50$, $c_2^R=-37.8$ (green-dotted), and 
(a)-(b): $a_2^L=-31$, $a_2^R=-29.5$; (c)-(d): $a_2^L=-10$, $a_2^R=-9.5$.
The other Ginzburg-Landau parameters are those for the left and the right 
material given in Table I.
} 
\end{figure}
%%%%%%%%%%%%%%%%%%%%%%%%%%%%%%%%%%%%%%%%%%%%%%%%%%%%%%%%%%%%%%%%%%%%%%%%
Specifically, the solutions for $u_z$, $e_1$, $e_3$ in the material left,
(right) 
from the interface occupying the region $z<0$ ($z>0$) are written as
\begin{eqnarray}
  \label{100}
   u_z^{L(R)} = b^{L(R)} +f^{+(-)} , \\
  \label{101}
  e_1^{L(R)} =e_{10}^{L(R)} +G^{+(-)} ,~~
  e_3^{L(R)} =e_{30}^{L(R)} -\frac{1}{\sqrt{3}} G^{+(-)} ,
\end{eqnarray}
where the superscript $L$ ($R$) indicates the value of the corresponding quantity
in the left (right) material. For this choice, the integration constants
in Eq. (\ref{31}) are
$C^{\pm} = \pm \frac{2}{\sqrt{\tilde{\mu}}} \tanh^{-1} \left( \frac{-\tilde{\lambda}}
   { 2 \sqrt{\tilde{\kappa} \tilde{\mu}} } \right)$,
so that $f$ and its derivatives vanish on either side of the heterostructure far 
from the interface, in accordance with our earlier assumptions.

In order to obtain solutions for $u_z$, $e_1$, $e_3$ that are globally valid 
over the whole bilayer structure, 
we impose the following (internal) boundary conditions at the interface
\begin{equation}
\label{33}
  u_z^L (z^\star) = u_z^R (z^\star), \qquad 
  \sigma_{zz}^L (z^\star) = \sigma_{zz}^R (z^\star) ,
\end{equation}
%%% (Note that the other two tractions on the interface are identically zero everywhere.)
where the stress component $\sigma_{zz}$ is obtained from $\Phi_z$, Eq. (\ref{21}), 
in the static limit,
and $z^\star$ is the location of the interface that is not necessarily at zero.
We thus distinguish between the positions of the actual interface, where the strains
exhibit significant variation, and the interface which is the natural boundary
of the two materials. The actual and the natural interfaces could be slightly displaced 
one another due to reconstruction of the interface, 
similarly to that observed in Ag(111)/Ru(0001) \cite{Ling2004}.
Eqs. (\ref{33}) can be satisfied for appropriate 
values of $z_0$ and $z^\star$ which can be obtained numerically.

The strains $e_1$ and $e_3$ along the $z-$direction, that is perpendicular to the 
interface, are shown in Fig. 4 for several combinations of $c_2$ and $a_2$.
The strain $e_1$ exhibits a minimum close to the interface, indicating relative 
lattice compression in that region.
Notably, compressionally strained layers at the PbTiO$_3$/SrTiO$_3$ interface,
corresponding to a reduced $c-$axis lattice parameter of the PbTiO$_3$ film
in the first few unit cells, have been experimentally observed \cite{Helvoort2005}.
This {\em interface reconstruction} is solely due to the elastic properties of
the materials of the bilayer. 
The dependence of $e_1$ on $z$ and $c_2$ is shown in the left panel of Fig. 5.
The corresponding dependence of the well's depth, $D$, and the constants
$z^\star$ and $z_0$, is shown in the right panel of Fig. 5. 
Thus, with decreasing $c_2$ (i.e., becoming more negative) $D$ increases, 
while $z_0$, which is a measure of the well's width, decreases.
Also, $z^\star$ decreases with decreasing $c_2$, so that the actual interface 
approaches the natural one at low temperatures.
%%%%%%%%%%%%%%%%%%%%%%%%%%%%%FIG.05%%%%%%%%%%%%%%%%%%%%%%%%%%%%%%%%%%%%%
\begin{figure}[t!]
\includegraphics[scale=0.46,angle=0]{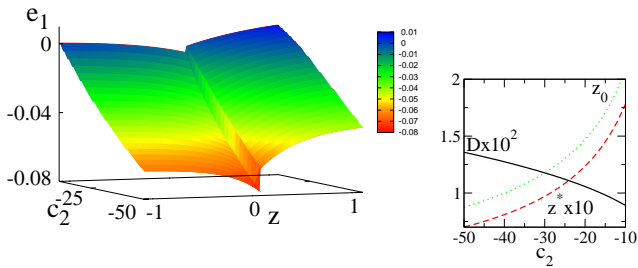}
\includegraphics[scale=0.26,angle=0]{IntHetero_Fig05b.eps}
\caption{
(color online)
Left panel: The strain $e_1$ as a function of the $z-$coordinate and the 
$c_2$ parameter. 
Right panel: The depth of the well $D$, and the numerically 
obtained constants $z_0$ and $z^\star$ as a function of $c_2$.
The other Ginzburg-Landau parameters are those for the left and the right 
material given in Table I.
}
\end{figure}
%%%%%%%%%%%%%%%%%%%%%%%%%%%%%%%%%%%%%%%%%%%%%%%%%%%%%%%%%%%%%%%%%%%%%%%%
 
The values of the Ginzburg-Landau parameters used in the calculations of the 
strains in Figs. 4 and 5 are given in Table I, calculated from the corresponding
elastic coefficients through Eqs. (\ref{9}). For the left ($L$) material, 
the second- and third-order constants are those reported for SrTiO$_3$ 
\cite{Bell1963,Beattie1971},
while the second-order constants for the right ($R$) material are those 
reported recently for LaTiO$_3$ \cite{Liu2011}. 
The values of the elastic coefficients, whose values are not reported in the
literature due to the lack of experimental and theoretical data,
are chosen to be reasonable for perovskites (Table I).
However,the obtained interfacial effects persists for a wide range of values
of the higher-order elastic coefficients. 
The strain gradient coefficients $d_1$ and $d_2$ are treated as phenomenological 
parameters to set the scale of the deformed region. Their values for the left and
the right material are chosen to be, respectively,  $d_1^L =d_2^L =6$ and 
$d_1^R =d_2^R =7$, in units of $10^{-7}~N$. For this choise, significant variations
in $e_1$ and $e_3$ occur within $\sim 1$ nm corresponding to $\sim$2-3 TMO layers.
The value of $c_2$ was taken to be negative for both materials,
in accordance with common practices of Ginzburg-Landau theory of phase transitions.
In Figs. 4 and 5, where $c_2$ and/or $a_2$ vary for both materials, a constant
ratio between the left and the right value was assumed, calculated from their
values given in Table I.  The parameter $a_2$ is again treated phenomenologically,
so that pairs of $a_2^L$ and $a_2^R$ with absolute values smaller that those given
in Table I (but with the same ratio) have been used.  
These values are also consistent with the principle $e_1 << e_2, e_3$ 
for all martensitic tranformations.

It has been reported that LaAlO$_3$ and LaTiO$_3$ follow the structure
of the SrTiO$_3$ substrate when the latter undergoes a CTT transition 
\cite{Kim2003,Wu2011}. 
The correlation of these structural changes and the electromagnetic 
properties in these 
systems may be empirically seen in the observation of an enhancement in
the interfacial charge carrier mobility and magnetization below $T_s$ 
\cite{Huijben2006,Kim2010,Christen2008,Ariando2011}.  
In TMO heterostructures undergoing a CTT transition,
significant lattice deformation occurs at the interface region due to lattice 
mismatch, that results in spontaneous strains.  
Despite the empirical evidence for the effects of interfacial lattice deformation
on the electromagnetic properties of TMO heterostructures, 
the relationship between the strain and the formation of a 2DEG remains 
largely unexplored.
In a Ginzurg-Landau approach that includes charge and/or magnetic degrees
of freedom, the dilatational strain $e_1$ couples linearly to the charge
density \cite{Bishop2003}. Then, the results of Fig. 4 reveal that $e_1$ 
serves as an effective potential well which may affect the charge
distribution throughout the heterostructure. 
In particular, the compressed interfacial region may attract and confine
electron charges.
The localized charges may contribute to the formation of a 2DEG,
a prerequisite for interfacial metallicity in TMO heterostructures.

%%%%----------Table1-------------------------------------------------
\begin{table}[h!]
  \centering  \label{table1}
        \begin{tabular}{||l|l|l||}  \hline  \hline
 Elastic   & Left  & Right       \\
 Const.    & Mater.& Mater.    \\ \hline \hline
 $C_{11}$  & 3.172 & 2.979  \\ \hline
 $C_{12}$  & 1.025 & 1.355  \\ \hline \hline
 $C_{111}$ &-50.0  &-47.5   \\ \hline
 $C_{112}$ &-4.0   &-3.8    \\ \hline
 $C_{123}$ &-3.0   &-2.85   \\ \hline \hline
 $C_{1111}$& 777.5 & 760.0  \\ \hline
 $C_{1112}$& 270.0 & 152.0  \\ \hline
 $C_{1122}$& 326.0 & 342.0  \\ \hline
 $C_{1123}$& 250.0 & 244.0  \\ \hline \hline
  \end{tabular}
        \begin{tabular}{||l|l|l||}  \hline  \hline
 GL     & Left  & Right       \\
 Coeff. & Mater.& Mater.    \\ \hline \hline
 $c_{1}$& 1.74  & 1.89   \\ \hline
 $c_{2}$& 4.29  & 2.25   \\ \hline \hline
 $a_{1}$&-3.00  &-2.80   \\ \hline
 $a_{2}$&-31.0  &-29.5   \\ \hline
 $a_{4}$& 25.4  & 24.1   \\ \hline \hline
 $b_{1}$& 48.7  & 42.9   \\ \hline
 $b_{2}$& 63.3  & 36.9   \\ \hline
 $b_{4}$& 225   & 393    \\ \hline \hline
 $b_{7}$&-35.4  &-40.3   \\ \hline \hline
  \end{tabular}
\caption{Second-, third-, and fourth-order elastic coefficients 
(in units of $10^{11} ~N/m^2$), and Ginzburg-Landau (GL) dimensionless 
coefficients for the left and the right material used in the calculations.
}
\end{table}

\section{Conclusions
}
We applied continuum elasticity to investigate theoretically the strain-state
of bilayer TMO heterostructures within a Landau theory,
and we have obtained simple approximate solutions for the fields $e_1$ and $e_3$.
Interface reconstruction may lead to electronic charge redistribution in the 
heterostructure, and particularly to electronic charge concentration in the 
interface region favoring the formation of a 2DEG.
The presence of a minimum in the dilatational strain field demonstrates
that possibility, linking thus the elastic to the electronic properties of TMOs.
Although such a reconstruction is a microscopic phenomenon involving significant 
changes of atomic arrangements at the interface \cite{Chakhalian2007}, 
it results in macroscopic changes of the unit cells that can be observed 
experimentally.
Those changes can be described, at least qualitatively, by the Ginzburg-Landau
theory, and their implications on the electron charge distribution of the bilayer
can be inferred from basic physical laws. 

\section*{Acknowledgments
}
This work was supported by the EURYI, MEXT-CT-2006-039047, 
and the National Research Foundation of Singapore.
We thank K. Rogdakis for useful discussions.

\appendix

\section{Nonlinear functions
}
The nonlinear functions $R_i$ ($i=x,y,z$) are given by
\begin{eqnarray}
 \label{a1}
  R_x=-a_4 \left[ +e_2 e_3 +\frac{1}{2\sqrt{3}} ( e_2^2 -e_3^2 ) \right]
    \nonumber \\
           +\frac{b_4}{2} (e_2^2 +e_3^2 )\left(+e_2 +\frac{1}{\sqrt{3}} e_3 \right)
    \nonumber \\
      +\frac{3 a_1}{2} e_1^2 
      +\frac{a_2}{2} 
     \left[ (e_2^2 +e_3^2) +e_1 \left(+e_2 +\frac{1}{\sqrt{3}} e_3 \right) \right]
                                \nonumber \\
      +b_1 e_1^3 +b_2 e_1 (e_2^2 +e_3^2)
      +\frac{b_2}{2} e_1^2 \left(+e_2 +\frac{1}{\sqrt{3}} e_3 \right)
    \nonumber \\
     +\frac{b_7}{2} \left[ e_3 ( e_3^2 -3 e_2^2 ) -3e_1 e_2 e_3 
                          +\frac{\sqrt{3}}{2} e_1 (e_3^2 -e_2^2) \right] 
\end{eqnarray}
\begin{eqnarray}
 \label{a2}
  R_y=
  -a_4 \left[-e_2 e_3 +\frac{1}{2\sqrt{3}} ( e_2^2 -e_3^2 ) \right]
    \nonumber \\
           +\frac{b_4}{2} (e_2^2 +e_3^2 )\left(-e_2 +\frac{1}{\sqrt{3}} e_3 \right)
    \nonumber \\
      +\frac{3 a_1}{2} e_1^2 
      +\frac{a_2}{2} 
     \left[ (e_2^2 +e_3^2) +e_1 \left(-e_2 +\frac{1}{\sqrt{3}} e_3 \right) \right]
                                 \nonumber \\
      +b_1 e_1^3 +b_2 e_1 (e_2^2 +e_3^2)
      +\frac{b_2}{2} e_1^2 \left(-e_2 +\frac{1}{\sqrt{3}} e_3 \right)
    \nonumber \\
     +\frac{b_7}{2} \left[ e_3 ( e_3^2 -3 e_2^2 ) +3e_1 e_2 e_3 
                          +\frac{\sqrt{3}}{2} e_1 (e_3^2 -e_2^2) \right]
\end{eqnarray}
\begin{eqnarray}
 \label{a3}
  R_z=-\frac{1}{\sqrt{3}} 
     \left[-a_4 (e_2^2 -e_3^2) +b_4 e_3 (e_2^2 +e_3^2 ) \right]
    \nonumber \\
   +\frac{3 a_1}{2} e_1^2 
   +\frac{a_2}{2} \left[ (e_2^2 +e_3^2) -\frac{2}{\sqrt{3}} e_1 e_3 \right]
    \nonumber \\
   +b_1 e_1^3 +b_2 \left[ e_1 (e_2^2 +e_3^2) -\frac{1}{\sqrt{3}} e_1^2 e_3 \right]
                                \nonumber \\
   +\frac{b_7}{2} \left[ e_3 ( e_3^2 -3 e_2^2 ) -\sqrt{3} e_1 ( e_3^2 -e_2^2 ) 
                  \right] 
\end{eqnarray}
It can be easily checked that for $e_2=0$ we have $R_x =R_y$.

\end{document}